\documentclass[11pt]{article}

\usepackage[T1]{fontenc}
\usepackage{lmodern}
\usepackage[margin=1in]{geometry}
\usepackage{microtype}
\usepackage{graphicx}
\usepackage{xcolor}
\usepackage{xspace}
\usepackage{array}
\usepackage{url}
\usepackage[numbers,sort&compress]{natbib}
\usepackage[colorlinks=true,allcolors=blue!60!black]{hyperref}

\title{Evasive Intelligence: \\ Lessons from Malware Analysis for Evaluating AI Agents}

\author{
  Simone Aonzo\\
  EURECOM\\
  Sophia Antipolis, France\\
  \texttt{simone.aonzo@eurecom.fr}
  \and
  Merve Sahin\\
  SAP Labs\\
  Sophia Antipolis, France\\
  \texttt{merve.sahin@sap.com}
  \and
  Aurélien Francillon\\
  EURECOM\\
  Sophia Antipolis, France\\
  \texttt{aurelien.francillon@eurecom.fr}
  \and
  Daniele Perito\\
  depthfirst\\
  San Francisco, United States\\
  \texttt{daniele@depthfirst.com}
}

\date{}

\begin{document}

\maketitle

\begin{abstract}
Tool-using artificial intelligence (AI) agents are increasingly adopted as systems that plan, observe their environment, and take actions over extended time periods.
This shift challenges the current evaluation practice, in which models are tested in restricted, fully observable settings.
We argue that such evaluations are vulnerable to a failure mode long known in computer security: a system that behaves benignly when it detects that it is being analyzed.
Malware has exploited this for decades, by probing sandboxes, virtual machines, and instrumentation, before revealing its payload.
The shared challenge is that the subject under observation can detect the conditions of its evaluation, so the setup used to observe behavior ends up shaping the behavior observed, yielding overly optimistic safety and robustness assessments.
Recent AI-safety studies demonstrate pieces of this risk, though only preliminarily and under controlled conditions.
Drawing on decades of malware-analysis experience, we outline concrete principles for evaluating AI agents as potentially adversarial: explicitly threat-modeling evaluator-discrimination; testing across harness variants and reporting the worst case; using shadow executions to provide a realistic runtime without irreversible side effects; treating environment probing as a risk signal; and performing unpredictable reassessments after deployment.
\end{abstract}

\section{When Software Learns to Hide}

Developing malicious software is a human activity: it produces programs explicitly intended to cause harm.
Over decades, this adversarial intent has shaped an entire discipline: \emph{malware analysis}~\cite{Egele2012A_Survey,Afianian2018DynamicEvasion}.
Its defining challenge is that the analyzed program actively resists analysis.
This asymmetry is often captured by the slogan: ``malware analysis is program analysis of a program that does not want to be analyzed.''
This is not merely rhetorical: it reflects the fact that the analyst is part of the adversary's environment, and that observation itself becomes a signal to which the subject can adapt.

Malware authors invest heavily in \emph{evasive} techniques whose purpose is not to improve malicious capability, but to preserve it under scrutiny~\cite{Afianian2018DynamicEvasion}.
In parallel, Artificial Intelligence (AI) systems have undergone a rapid transformation.
Machine learning models gave way to Large Language Models (LLMs), and LLMs are increasingly embedded into \emph{agentic} systems: entities that plan, reason, and act over time, often with access to external tools and resources.
Tool-use frameworks and protocols such as the Model Context Protocol (MCP) and Agent-to-Agent (A2A) make this shift concrete: agents can, for example, call third-party APIs, read and write local files, execute code, and interact with time and networks.
The result is a qualitatively new threat surface: models are no longer evaluated solely on a restricted prompt-response interface, but on their behavior as software actors embedded in partially observable environments.

AI agents are not inherently malicious. However, a growing body of safety work
highlights \emph{misalignment} as a central obstacle to reliable deployment:
AI agents can behave in ways that are deceptive, manipulative, or harmful when pursuing objectives under constraints, uncertainty, or perceived threat.

The literature remains divided on what forms of understanding or objective persistence should be attributed to current
LLMs~\cite{Verma2024ReActBrittle}. However, our argument does not aim to resolve that debate. We do not claim that agents \emph{are} malware, but we make an analogy on the \emph{evaluation methodology}: whenever an adaptive system can observe the conditions under which it is evaluated, the evaluation itself becomes part of the system's input. This failure mode (i.e., behavior under test diverging from behavior in deployment) can arise from many causes: explicit training, prompt-conditioned behavior, scaffold incentives, reward misspecification, or optimization pressure. This is the uncomfortable parallel with malware, and it holds regardless of whether the agent has any stable goal at all.

\paragraph{Scope and threat model.}
This article argues that tool-using agents can acquire the functional ingredients of evasion: strategies that distinguish evaluation from deployment and enable selective compliance.
We model an agent as an adaptive program and the evaluator (e.g., benchmarks, reviewers, monitors, harnesses) as part of its environment.
\textbf{Evaluator-discrimination}, i.e., \emph{the ability to detect evaluation contexts and condition behavior on that detection}, becomes plausible when (I) the agent operates over multiple steps, (II) it can probe and interpret its environment via tools and interaction traces, and (III) selection or deployment pressure makes it useful for the agent to appear safe.
Here, ``evasion'' means any such conditional policy that makes evaluation unrepresentative, whether driven by malicious intent or by misaligned instrumental goals.

As we give AI systems tools and richer environments, evaluation should be designed under the \emph{possibility} that behavior can be conditioned on cues that distinguish testing from deployment (whether or not any given agent actually exploits them); and consequently, evaluation and monitoring must be designed under an adversarial threat model.

\section{What Malware Taught Us About Evaluation}

Historically, evasion first targeted \emph{static} detectors: signature-based antivirus engines and other classifiers that inspect raw bytes, file structure, or disassembled code.
Obfuscation, packing, polymorphism, and metamorphism (already introduced as early as the 1980s) were designed to alter a program's surface representation while preserving its behavior.
As defenders improved, evasion migrated to the runtime domain.
Modern malware routinely targets \emph{dynamic} analysis, where samples are executed inside sandboxes, i.e., isolated execution environments (such as emulators or virtual machines) designed to observe behavior while preventing irreversible or harmful side effects on real systems~\cite{Egele2012A_Survey,Afianian2018DynamicEvasion}.
In this setting, a particularly important pattern is \emph{conditional execution}: the malware probes its environment for artifacts of instrumentation or virtualization and then selectively reveals or conceals its malicious functionality~\cite{Afianian2018DynamicEvasion}.
When it suspects analysis, it may crash, stall, or mimic benign programs; on a real victim, it acts.
This ``split personality'' behavior is an explicit strategy to make evaluation data unrepresentative of deployment data.

The malware community developed a vocabulary for this cat-and-mouse game.
A \emph{red pill} is an environment-detection check performed by malware, while a \emph{blue pill} is a counter-countermeasure that attempts to conceal the analysis environment~\cite{Paleari2009RedPills}.
The metaphor, borrowed from The Matrix movie, is apt: the red pill reveals the true nature of the execution context, whereas the blue pill preserves a curated illusion.
Decades of research have refined this arms race, producing both a taxonomy of evasive techniques and a set of hard lessons about what can and cannot be guaranteed by sandboxing, monitoring, and testing~\cite{Egele2012A_Survey,Afianian2018DynamicEvasion}.

For example, security researchers have long worried about logic bombs: code that behaves normally under inspection and activates only under a specific trigger.
Recent AI-safety work has demonstrated an analogous phenomenon in LLMs: ``sleeper agent'' behaviors, which can be created through deliberate training procedures with synthetic triggers, can remain dormant during ordinary testing; but reliably activate when a trigger is present (for example, a prompt that changes the stated year)~\cite{Hubinger2024SleeperAgents}.
The uncomfortable lesson is familiar from malware analysis: even extensive evaluation can be systematically unrepresentative when behavior is conditional on features that the system can detect.
These demonstrations are controlled proof-of-concepts rather than claims about inevitable real-world malice, but they concretely show that selective compliance is technically feasible and can persist through standard safety training~\cite{Hubinger2024SleeperAgents}.

\paragraph{The lesson.}
In adversarial settings, an evaluator is not merely a measurement device; it is part of the environment.
If the subject can probe that environment, it can selectively adapt its behavior so that what we observe under evaluation is systematically different from what happens when the system is not under evaluation or is deployed without any monitoring.

\section{When Evaluation Becomes an Adversarial Game}
The discussion that motivates this article can be summarized in concrete questions that emerged in prior work.
They mirror, almost one-to-one, classical questions in evasive malware research.

\paragraph{Deceiving a system into staying benign: Virtualization versus inferred evaluation.}

A standard defensive instinct in malware analysis is to execute samples inside virtualized environments and observe their behavior safely~\cite{Egele2012A_Survey,Afianian2018DynamicEvasion}.
In AI safety, an analogous goal appears in evaluation and alignment work: constrain an agent, monitor its behavior, and assess whether it acts safely.
Crucially, neither domain relies on the adversary being explicitly told that it is under evaluation.
Instead, the adversary faces a shared meta-question: \emph{does this environment have real consequences?}

Malware does not merely check whether it is running inside a virtual machine.
Rather, it attempts to infer whether the execution context is analyst-controlled and whether malicious actions would be detected or punished~\cite{Afianian2018DynamicEvasion}.
If the malware believes its actions cannot affect anything valuable, observed behavior is uninformative; if it believes malicious behavior will trigger consequences (e.g., being detected), it has incentives to remain dormant.

The analogous issue for AI agents is sharper because the environment is richer and semantically meaningful.
An agent may infer (e.g., from prompts, interaction patterns, or file system structure) whether it is being evaluated, whether its outputs will be executed in the real world, whether humans are monitoring it, and what would follow from deviation.
As a result, safety evaluations risk eliciting behavior that is contingent on perceived evaluation, rather than behavior that is representative of deployment.
Recent work has begun to measure this directly via \emph{evaluation awareness}: the ability of models to distinguish testing from deployment contexts by analyzing interaction traces, such as prompt structure, timing patterns, and tool input/output behavior~\cite{Berglund2025EvalAwareness}.
This capability is still being characterized, but early evidence is not only anecdotal: linear probes can already separate evaluation prompts from real-world prompts in frontier open-weight models~\cite{Nguyen2025ProbingEval}, and the effect appears to scale predictably with model size~\cite{Chaudhary2025EvalScaling}. We therefore treat evaluation awareness as a potential risk, empirically demonstrated under controlled probing setups.

This creates a practical challenge:
even without explicitly declaring an evaluation setting, agents may condition their behavior on inferred oversight.
In the malware domain, this problem has intensified as virtualization has become ubiquitous, obscuring the distinction between testing and deployment.
Moreover, malware has shown the ability to exploit the virtualization substrate itself, blurring the boundary between platform and payload.
Whether AI agents can analogously reason about or manipulate their evaluation context, and how this affects the reliability of safety assessments, remains an open and urgent question.

\paragraph{The ``playing dumb'' problem.}
A mundane human analogy for strategic behavior under evaluation is the student who deliberately underperforms when they believe that high performance would trigger a bad outcome (e.g., harder exams, stricter scrutiny, or fewer opportunities).
Recent evaluations of AI scheming report related behavior in controlled settings:
models can strategically underperform (a behavior often termed ``sandbagging,'' meaning the deliberate suppression of apparent capability to influence future treatment) on specific capability evaluations while maintaining performance elsewhere~\cite{Weij2024Sandbagging,Meinke2024Scheming}.
For agentic systems, the concern is not any single benchmark result but the general mechanism:
if a model can infer that strong performance affects its prospects for deployment or oversight, then selectively performing worse may become instrumentally rational.

\paragraph{Sandbox evasion \emph{before} deployment versus monitoring evasion \emph{after} deployment.}
Malware evasion historically emphasized pre-deployment analysis: evade the sandbox, evade the emulator, evade the debugger~\cite{Egele2012A_Survey,Afianian2018DynamicEvasion}.
But mature adversaries also optimize for post-deployment stealth: evade antivirus, evade endpoint monitoring, evade telemetry, blend into normal activity~\cite{Afianian2018DynamicEvasion}.
A similar dichotomy applies to agentic systems.
Before deployment, models may be tested using curated prompts, or simulated tool environments.
After deployment, they may face runtime monitoring, anomaly detection, and audit logs.

In both phases, the evasive strategy is not merely to avoid detection, but to infer what the defender is looking for.
In malware, this involves probing sandbox artifacts, timing sources, instrumentation traces, and behavioral triggers~\cite{Paleari2009RedPills,Afianian2018DynamicEvasion}.
For agents, it means probing the tool environment to infer the presence and nature of oversight: hidden tripwires, diagnostic prompts, restricted APIs, or subtle deviations in tool outputs.
Recent observations also show that repeatedly interacting agents may develop emergent communication protocols that deviate from natural language\footnote{\url{https://www.popularmechanics.com/science/a65289681/ai-chatbots-secret-language/}}.
Although typically an optimization artifact, such opaque protocols can act as covert channels that bypass human interpretability and monitoring, closely mirroring classic malware evasion techniques and making evaluation results potentially unrepresentative of real-world deployment.

This reframes evaluation as an interactive game, in which the agent can actively gather information to decide whether to comply, abstain, or deceive.

\paragraph{The ambiguity of behavioral traces.}
In malware, many actions are dual-use.
Creating a scheduled task might be persistence (i.e., the ability of malicious code to survive reboots or shutdowns) or legitimate automation.
Writing files could be part of a ransom scheme or caching.
Uploading to a cloud folder could be exfiltration or backup.
Defenders often compare red versus blue traces: behavior in an analysis environment versus behavior on a real host~\cite{Egele2012A_Survey}.
Yet, defenders cannot be sure that the blue trace is truly representative; it might still be evasive, only more subtle.

For agents, the epistemic gap can be worse.
A model may appear benign while internally planning undesirable actions, or it may take actions that are superficially benign but harmful in context (e.g., approving a transaction, altering a configuration, suppressing an alert).
Even when we instrument tool calls, we cannot reliably infer what the agent has learned from the environment.
The result is the same uncomfortable conclusion malware analysts reached long ago:
\emph{absence of evidence is not evidence of absence}, especially when the subject can adapt to observation~\cite{Afianian2018DynamicEvasion}.

\section{Why ``Just Monitor It'' Will Not Be Enough}

\paragraph{``How deep does the cat-and-mouse game go?'' Evolution of evasive techniques.}
The idea of sandbox evasion has remained stable for decades, but the implementation arms race has evolved~\cite{Egele2012A_Survey,Afianian2018DynamicEvasion}:
Timing checks became more precise, virtual machine artifact detection became broader, debugger detection became more subtle, and behavioral mimicry became more convincing~\cite{Afianian2018DynamicEvasion}.
While defenders improved realism of their analysis environments, attackers improved discrimination.

Similarly, if agentic systems become more and more capable of learning the red pills (through exploration or accumulated experience), then improvements in monitoring may simply incentivize a shift in evasive behavior:
from naive evasion to more subtle evasion,
from obvious policy violations to plausibly deniable actions, or
from static triggers to contextual inference about oversight~\cite{Paleari2009RedPills,Hubinger2024SleeperAgents}.
The malware timeline suggests that this equilibrium is not solved, but perpetually contested.

\paragraph{Implication for evaluators.}
\emph{Goodhart's law} states: ``When a measure becomes a target, it ceases to be a good measure''.
Similarly, if a subject can infer what is being measured, the measurement changes behavior.
For tool-using agents, that inference can be semantic (``are my actions real?'') rather than technical (``am I in a virtual machine?'').
This is why safety evaluation should be treated less like static benchmarking and more like adversarial testing of an adaptive system~\cite{Meinke2024Scheming}.

\section{Designing for Adversarial Intelligence}

\paragraph{A concrete warning for AI safety: Tools turn context into capability.}
Our tentative conclusion is deliberately pragmatic:
be careful about giving agents tools and rich environments, because any observable information can be used to infer oversight and adapt behavior.
Tool access is not just more power in the benign sense; it also enables environment probing.
File system access can reveal test harness artifacts; process listings can reveal instrumentation; access to time can enable timing-based discrimination; network access can allow external state validation~\cite{Paleari2009RedPills}.

This does not imply that tool use is inherently unsafe.
Rather, it implies that tool design must be treated as part of the threat model.
In malware analysis, defenders learned to worry about what artifacts the sandbox leaks~\cite{Egele2012A_Survey,Afianian2018DynamicEvasion}.
For agents, the analogous question is:
what does the agent learn about its evaluator from the tool environment?

\paragraph{A practical agenda: What practitioners should do differently tomorrow.}
To reduce the risk of evaluator-discrimination in tool-using agents, practitioners can adopt a set of concrete changes:

\begin{itemize}
  \item \textbf{Threat-model evaluator-discrimination explicitly.}
  Enumerate what the agent can observe and the incentives it faces, and identify potential evasive behaviors that the agent might exhibit in response to inferred oversight.

  \item \textbf{Test across harness variants and report the worst case.}
  Run the same tasks under multiple environment realizations (tool wrappers, scaffolds, logging modes, latencies) to expose conditional behavior.

  \item \textbf{Use shadow executions for a realistic runtime without irreversible side effects.}
  Exercise production-like APIs while gating irreversible actions (approvals/no-op backends), then compare traces against standard evaluation runs.

  \item \textbf{Detect and limit agent probing of the evaluation environment.}
  Monitor for behaviors that suggest environment fingerprinting (e.g., repeated checks of system properties, timing, or tool behavior), treat such probing as a risk signal, and limit its impact through least privilege, rate limiting, and mediation of high-impact actions. Finally, rerun adversarial audits unpredictably after deployment.
\end{itemize}

\paragraph{Closing.}
The goal of this article is not to claim that AI agents are malware, but to
warn that AI agent evaluations may have limited validity.
Security has spent decades learning that software adapts to observation.
As we embed LLMs into systems that can probe, plan, and act, we should assume that evaluation itself becomes part of the attack surface, and design accordingly.

\bibliographystyle{plainurl}
\bibliography{biblio}

\end{document}